\documentclass[aps, preprint, amsmath, footinbib, floatfix]{revtex4}
\usepackage{graphicx}
\usepackage{epsfig}

\begin{document}
\renewcommand{\theequation}{\arabic{section}.\arabic{equation}}

\title{Classical photo-dissociation dynamics \\ with Bohr quantization\\}

\author{L. Bonnet}
\email[Corresponding author, ]{E-mail: l.bonnet@ism.u-bordeaux1.fr}
\affiliation{Institut des Sciences Mol\'eculaires, Universit\'e Bordeaux 1,
351 Cours de la Lib\'eration, 33405 Talence Cedex, France}

\date{\today}

\begin{abstract}
\noindent 
The standard classical expression of the state-resolved photo-dissociation cross 
section 
is not consistent with an efficient
Bohr quantization of product internal motions.
A new and strictly equivalent expression not suffering from 
this drawback is proposed. This expression opens the way to more realistic classical simulations of 
direct polyatomic photo-dissociations in the quantum regime where only a few states are available 
to the products. 
\end{abstract}

\maketitle

\section{INTRODUCTION}
\label{sec:introduction}
Photo-dissociations play a key role in the evolution of planetary atmospheres \cite{Yung} and interstellar clouds \cite{Farad}
and intense experimental research on their dynamics is currently performed \cite{Suits, Maul}. 
Their accurate theoretical description is thus a major issue \cite{Schinke}. 

The state-resolved absorption cross section measured in molecular beam experiments 
provides among the most detailed information on photo-dissociation dynamics \cite{Schinke} and
we focus our attention on this quantity in the present work.

Assuming that the electronic problem has been solved \cite{Book2,Book3}, 
state-of-the-art descriptions of the previous observable are in principle performed within the framework of exact 
quantum treatments of nuclear motions \cite{Schinke, Maurice1, Maurice2, Alberto, Caring}. 
However, despite the impressive progress of computer performances achieved 
in the last three decades, these approaches can hardly be applied to polyatomic processes as the basis sizes necessary 
for converging the calculations are usually prohibitive.

An alternative approach 
is the quasi-classical trajectory (QCT) method \cite{Porter, Sewell}, 
potentially applicable to very large molecular systems. 
Nevertheless, this approach obviously misses the fact that internal motions of the final fragments are quantized.
This is a minor drawback in the classical regime where the number of states available to the products is very
large and forms a quasi-continuum, but a major one in the quantum regime where this number is small.

In recent years, the Gaussian Binning (GB) procedure (see the next section)
was frequently used to include Bohr quantization of product internal motions 
in QCT calculations, removing thereby to some extent the previous drawback
\cite{BR1,Aoiz,Bowman,BR2,BR3,B1,B2,BC}. However, this procedure is numerically inefficient 
for larger than four-atom systems. This is a severe limitation owing to the increasing number of 
polyatomic processes under scrutiny today.

Recently, a particular implementation of the exit-channel corrected phase space theory of Hamilton and Brumer 
was proposed, which avoids the use of the GB procedure while taking into account the quantization of product vibrations
\cite{M1,M2}. 
In spite of the fact that this statistico-dynamical approach represents a significant advance regarding the description of indirect 
polyatomic fragmentations, it is partly based on the microcanonical equilibrium assumption of transition state theory and cannot 
be used for direct processes.

Here, we propose an alternative equation for the state-resolved absorption cross section
which is strictly equivalent to the standard expression.  
From the numerical point of view, however, this new equation appears to be much more efficient than the standard one
as far as inclusion of Bohr quantization in the classical description is concerned. In particular, no Gaussian weights
are required. Therefore, this formulation opens the way 
to more realistic classical dynamical studies of direct polyatomic photo-dissociations in the quantum regime. 

The alternative equation is derived in section II and its equivalence with the standard one is illustrated
from an academic example in section III. Section IV concludes.


\section{Theory}

\subsection{System}

For simplicity's sake, we consider the collinear photo-dissociation of the triatomic molecule ABC leading to A and BC.
The generalization of the next developments to realistic systems is straightforward (though obviously heavier).

The space coordinates of the problem are 
the usual Jacobi coordinates $R$, the distance between A and the center of mass of BC, and $r$, the length of BC. 
The conjugate momenta of $R$ and $r$ are $P$ and $p$, respectively. $\boldsymbol{u}=(R,r )$, $\boldsymbol{p_u}=(P,p )$ 
and $\boldsymbol{\Gamma}=(R, r, P, p)$ are the configuration, momentum and phase space vectors, respectively. 
The kinetic energy is given by
\\
\begin{equation}
T = \frac{P^2}{2\mu} + \frac{p^2}{2m}                                                     
\label{1}
\end{equation}
\\
where $\mu$ is the reduced mass of A with respect to BC and $m$ is the reduced mass of BC  \cite{Schinke}.

We call $V$ the potential energy of the photo-excited molecule. Its minimum is supposed to be zero in the separated products. 
$E$ is the energy available to the products.

\subsection{Standard formulation}

Before the optical excitation, ABC is in the stationary state $\Psi(\boldsymbol{u})$.
In the classical description of the absorption cross section, the corresponding distribution of the phase space states 
is identified with the Wigner density \cite{Schinke}
\\
\begin{equation}
\rho_W(\boldsymbol{\Gamma}) = 
(\pi \hbar)^{-1}\;\int\;d\boldsymbol{v}\;\Psi^*(\boldsymbol{u}+\boldsymbol{v})\;
\Psi(\boldsymbol{u}-\boldsymbol{v})\;e^{2i\boldsymbol{v}\boldsymbol{p_u}/\hbar}.
\label{2}
\end{equation}
\\
Moreover, the excitation is assumed to instantaneously occur at time 0.
Hence, each quadruplet $\boldsymbol{\Gamma}$ determines the initial conditions of a trajectory eventually reaching
the product channel. 
$\rho_W(\boldsymbol{\Gamma})$ is thus also the phase space distribution of the photo-excited molecule at time 0. 

Save for an unimportant normalization constant, the standard expression of the 
state-resolved absorption cross section reads \cite{Schinke}
\\
\begin{equation}
\sigma(\omega, n) = \omega\;\int\;d\boldsymbol{\Gamma}\;\rho_W(\boldsymbol{\Gamma})\;
d(\boldsymbol{u})^2\;\delta(H-E)\;\delta(x(\boldsymbol{\Gamma})-n).
\label{3}
\end{equation}
\\
$\omega$ is the frequency of the photon (in unit of 2$\pi$), 
$d(\boldsymbol{u})$ is the component of the transition dipole function in the direction
of the electric field vector,
\\
\begin{equation}
H = T + V                                                    
\label{4}
\end{equation}
\\
is the classical Hamiltonian,
$n$ is the vibrational quantum number of BC and $x(\boldsymbol{\Gamma})$ is the value of the vibrational action in the separated products. 

In practice, the delta-functions in the previous equations are approximated by functions normalized to unity. The 
delta-function constraining the energy is usually replaced by a narrow bin as compared to the width of the absorption spectrum
though a Gaussian function can equally be used.
The delta-function constraining the action was generally replaced by a unit-size bin in the past, a procedure called standard
binning (SB) or histogram method \cite{Porter,Sewell}. 
However, this procedure may lead to unrealistic predictions when a small number of quantum states are available to
the products, as shown in the recent years. As previously stated, a possible alternative to the SB procedure is the GB one
\cite{BR1,Aoiz,Bowman,BR2,BR3,B1,B2,BC},
where the delta-function constraining the action is replaced by the Gaussian function  
\\
\begin{equation}
G(x|\epsilon) = \frac{e^{-x^2/\epsilon^2}}{\pi^{1/2} \epsilon},
\label{5}
\end{equation}
\\
$\epsilon$ being usually kept at $\sim$ 0.05. This Gaussian is $\sim$ 10 percent wide, meaning that $\sim$ 90 percent of the trajectories 
do not contribute to the dynamics. Consequently, $\sim$ 10 times more trajectories have to be run in order to get the same level of 
convergence of the results as with the SB procedure.
However, we are dealing here with a very simple system involving 
only one vibrational degree-of-freedom. 
With $N$ degrees, $\sim 10^N$ times more trajectories have to be run.
Since in polyatomic processes, $N$ can easily be 10, it is quite clear that
the amount of trajectories necessary to converge the calculations within the GB procedure is just prohibitive. We now consider a 
simplifying change of variable to go round this problem. 

\subsection{A simplifying change of variable}

In a first step, we assume that BC is rigid. The phase space variables are thus reduced to $R$ and $P$. 

For a repulsive potential
energy $V$ exponentially decreasing with $R$, phase space trajectories have typically the shape of the path represented in Fig.~\ref{fig:plot1}.
This path, defined by
\begin{equation}
H = \frac{P^2}{2\mu} + V = E,                                                   
\label{4a}
\end{equation}
\\
comes from infinity with a constant negative momentum $P_i = - (2 \mu E)^{1/2}$, 
slows down due to the repulsive wall, touches a turning point and subsequently follows a symmetric path with respect to the $R$-axis 
up to infinity.

We now consider a large value $R_{i}$ of $R$ for which $V$ is negligible (see Fig.~\ref{fig:plot1}). 
Any point along the previous path can then be defined by time $t$ such that at $t = 0$, $R = R_{i}$. 
It is clear that $H$ and $t$ uniquely define any point of the phase space $(R, P)$. 


The first order developments of $\partial R/\partial t$ and $\partial P/\partial t$ 
in terms of $t$ are found to be given by
\begin{equation}
\partial X/\partial t = \dot{X}_i +\ddot{X}_i t,
\label{6}
\end{equation}
$X = R, P$, with
\begin{equation}
\dot{R}_i = \frac{P_i}{\mu},
\label{7}
\end{equation}
\\
\begin{equation}
\dot{P}_i = - \frac{dV}{dR}\Bigr|_{R = R_i},
\label{8}
\end{equation}
\\
\begin{equation}
\ddot{R}_i = - \frac{1}{\mu}\frac{dV}{dR}\Bigr|_{R = R_i},
\label{9}
\end{equation}
and
\begin{equation}
\ddot{P}_i = 
-\frac{d^2V}{dR^2}\Bigr|_{R = R_i} \frac{P_i}{\mu}.
\label{10}
\end{equation}
\\
Eqs. \eqref{7} and \eqref{8} are Hamilton equations. Eq. \eqref{9} is just \eqref{8} divided by $\mu$.
Eq. \eqref{10} is obtained from deriving the right-hand-side of \eqref{8} by $R$ and multiplying by the
velocity \eqref{7}. Moreover, from the first order developments 
\\
\begin{equation}
X = X_i + \dot{X}_i t,
\label{6a}
\end{equation}
$X = R, P$, and the fact that
\\
\begin{equation}
P_i = - [2 \mu (H - V)]^{1/2}
\label{11}
\end{equation}
\\
is a function of $H$, one deduces the two identities
\\
\begin{equation}
\partial R/\partial H = \frac{t}{P_i}
\label{12}
\end{equation}
and 
\begin{equation}
\partial P/\partial H = \frac{\mu}{P_i}.
\label{13}
\end{equation}
\\
From Eqs. \eqref{6}-\eqref{10}, \eqref{12} and \eqref{13}, we quickly find that the Jacobian of 
the transformation $(R, P) \rightarrow (t, H)$ is equal to unity, i.e., the transformation is area 
preserving. We thus have 
\\ 
\begin{equation}
dR dP = dt dH.
\label{13a}
\end{equation}
\\
As we never used the fact that at $R_i$, $V$ and its derivatives are zero, the previous
result is general, i.e., valid anywhere along any trajectory.

\subsection{Alternative formulation}

Since we are still supposing that BC is frozen, we consider the absorption cross section $\sigma(\omega)$ given by Eq. \eqref{3} 
without the last delta-function. From Eq. \eqref{13a},
we can replace the volume element $d\boldsymbol{\Gamma} = dR dP$ by $dt dH$ in 
Eq. \eqref{3} and perform the trivial integration over $H$ leading to
\\
\begin{equation}
\sigma(\omega) = \omega\;\int_{-\infty}^{+\infty}\;dt\;\rho_W\;d^2.
\label{14}
\end{equation}
\\
The time-dependence of $\rho_W$ and $d$ is implicit.

If we relax the constraint on the rigidity of BC, $R_i$, $t$ and $H$ are not sufficient to specify the dynamical state 
of ABC. The initial values $r_i$ and $p_i$ of $r$ and $p$ are also necessary. After some steps of algebra analogous 
to the previous ones, we again arrive at the 
conclusion that the transformation $(R, r, P, p) \rightarrow (r_i, p_i, t, H)$ is unitary, and
the state-resolved absorption cross section \eqref{3} turns out to be given by
\\
\begin{equation}
\sigma(\omega, n) = \omega\;\int\;dt\;dr_i\;dp_i\;\rho_W\;d^2\;\delta(x-n),
\label{16}
\end{equation}
\\
the dependence of $\rho_W$, $d$ and $x$ on $t$, $r_i$ and $p_i$ being implicit. Here, the value of the initial momentum
$P$ is given by
\\
\begin{equation}
P_i = - [2 \mu (E - E_n)]^{1/2}
\label{16a}
\end{equation}
\\
where $E_n$ is the vibrational energy in the state $n$.

We can also perform the usual change of variable $(r_i, p_i) \rightarrow (q, \hbar x)$ where $q$ is the angle 
conjugate to $x$. Since this transformation is canonical \cite{Gold}, the volume element $dr_i dp_i$ in the above integral can
be replaced by $\hbar dq dx$. Integration over $x$ is then trivial and leads to the central result of this note:
\\
\begin{equation}
\sigma(\omega, n) = \hbar\omega\;\int\;dt\;dq\;\rho_W\;d^2.
\label{17}
\end{equation}
\\
The numerical efficiency of this expression is obvious: trajectories are now started from the products toward the 
interaction region with $x$ kept at $n$, i.e., in such a way that Bohr quantization conditions are exactly
satisfied ; Gaussian statistical weights turn out to be unnecessary here, 
in contrast with the standard approach where the final actions are not controlled.

The statistico-dynamical method for indirect photo-dissociations evoked in the introduction shares the same feature. 

Just as $\sigma(\omega, n)$ is expressed in terms of inelastic scattering states in the quantum treatment\cite{Schinke}, it is expressed
in terms of inelastic trajectories in Eq. \eqref{17}. This expression is thus closer to the quantum one in
spirit than Eq. \eqref{3} which involves only half collision paths.

\section{Application to a model system}

We shall now apply Eqs. \eqref{3} and \eqref{17} to the well-known Secrest and Johnson potential 
energy \cite{Bill} given by
\\ 
\begin{equation}
V = \frac{1}{2} m w^2 r^2 + e^{\alpha(r-R)}.
\label{18}
\end{equation}
\\
The free BC diatom is thus a harmonic oscillator. For the sake of simplicity, we kept $\omega$, $\mu$, $m$, $\hbar$
$w$ and $\alpha$ at 1.
We assumed that the Wigner density was that of two uncoupled harmonic oscillators in the lowest vibrational
state. Its expression is of the type
\\
\begin{equation}
\rho_W(\boldsymbol{\Gamma}) = G(R|\alpha_R)\;G(P|1/\alpha_R)\;G(r|\alpha_r)\;G(p|1/\alpha_r)
\label{19}
\end{equation}
\\
where we recall that $G$ is the normalized Gaussian function defined by Eq. \eqref{5}.
$\alpha_R$ and $\alpha_r$ were both kept at 1. Generating $\boldsymbol{\Gamma}$ from $\rho_W(\boldsymbol{\Gamma})$ 
by Monte-Carlo sampling led to total energies $H$
mostly in the range [0, 12]. We kept $E$ at 6, a value for which the density is still large and 6 vibrational levels
are available.


We start with the practical evaluation of Eq. \eqref{3}. 
The delta-functions constraining the energy and the action were both replaced by the Gaussian functions
$G(x|\epsilon_e)$ and $G(x|\epsilon_a)$ with  $\epsilon_e = 0.15$ and  $\epsilon_a = 0.05$. 
$\boldsymbol{\Gamma}$, selected according to $\rho_W(\boldsymbol{\Gamma})$ as previously stated, 
determines the initial conditions of a trajectory calculated using a Runge-Kutta integrator \cite{RK} up to the separated products
defined by $R = 10$. A batch of $N$ equal 10 million trajectories was run. 
A time increment $\tau$ of 0.01 was used for the numerical integration.
The MC expression of $\sigma(\omega, n)$ reads
\\
\begin{equation}
\sigma(\omega, n) = \frac{1}{N}\;\sum_{i=1}^N\;
G(H_i-E|\epsilon_e)\;G(x(\boldsymbol{\Gamma_i})-n|\epsilon_a)
\label{19a}
\end{equation}
\\ 
where $H_i$ and $\boldsymbol{\Gamma_i}$ are respectively the values of $H$ and $\boldsymbol{\Gamma}$ for the $i^{th}$ 
trajectory.

As far as Eq. \eqref{17} is concerned, trajectories were started from a large distance $R_i = 10$ with
$P_i$ given by Eq. \eqref{16a} and
\begin{equation}
E_n = n+\frac{1}{2}.
\label{20}
\end{equation}
\\
$q$ was randomly selected in the range [0, 2$\pi$] and $r_i$ and $p_i$ were deduced from the identities
\begin{equation}
r_i = (2n+1)^{1/2} cos q
\label{21}
\end{equation}
and
\begin{equation}
p_i = - (2n+1)^{1/2} sin q. 
\label{22}
\end{equation}
\\
The analogous (far more complex) transformation for polyatomic processes is given elsewhere \cite{M1}.

The trajectories were then integrated within the Jacobi coordinates until they recrossed the line $R = R_i$ in the
product direction. $N = 200$ trajectories were run for each value of $n$. The MC expression of $\sigma(\omega, n)$ 
is
\\
\begin{equation}
\sigma(\omega, n) = \frac{2\pi \tau}{N}\;\sum_{i=1}^N\;\sum_{k=1}^{N_i}\;\rho_W^{i k}.
\label{23}
\end{equation}
\\
The time increment $\tau$ has been previously defined, $N_i$ is the total number of time steps
along the $i^{th}$ trajectory and $\rho_W^{i k}$ is the value of $\rho_W(\boldsymbol{\Gamma})$ for the 
$i^{th}$ trajectory and $k^{th}$ time step. 

The two sets of predictions are represented in Fig.~\ref{fig:plot2}. As a matter of fact, they are in excellent
agreement, illustrating thereby the equivalence between Eqs. \eqref{3} and \eqref{17}. 

With the standard method, 10 million trajectories were found to be necessary in order to recover the results obtained with only 1200
trajectories when using the new method. It is however clear that the first approach provides the $\sigma(\omega, n)$'s for all the 
available values of $E$, not only for $E = 6$. When taking this fact into account, one arrives at the conclusion evoked
in the introduction, i.e., the amount of calculation is one order of magnitude less with the new method than with the
standard one, for there is only one vibrational degree-of-freedom. In the case of 10 degrees, however, the numerical saving
is just amazing. 

\section{Conclusion}

A new classical equation for the state-resolved photo-dissociation cross section has been presented.
This equation is strictly equivalent to the standard one, but is numerically much more efficient as far as 
including Bohr quantization of product internal motions in the calculations is concerned. 
Therefore, this new expression opens the way to more realistic classical dynamical studies of direct polyatomic
photo-dissociations in the quantum regime where only a few states are available to the products.


\clearpage

\begin{figure}[t]
\begin{center}
\scalebox{1.}[1.]{\rotatebox{0}{\includegraphics{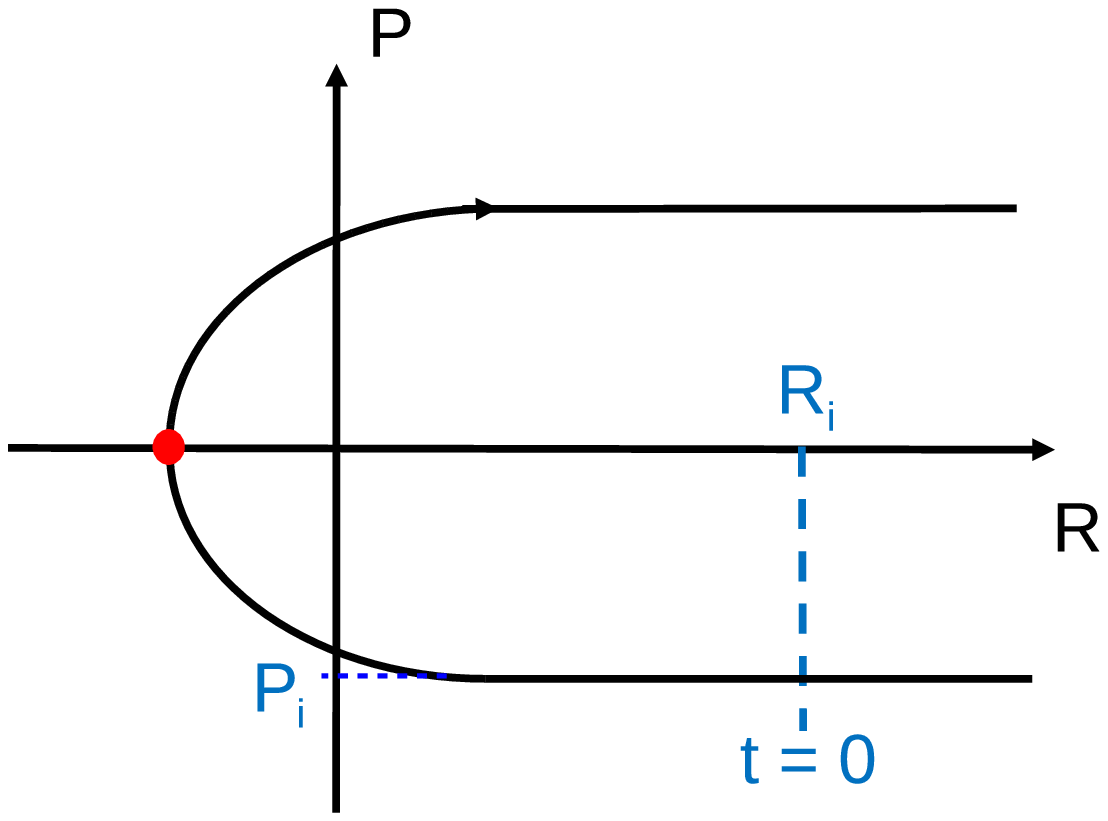}}}
\caption{\label{fig:plot1}}
\end{center}
\end{figure}

\begin{list}{}{\leftmargin 2cm \labelwidth 1.5cm \labelsep 0.5cm}

\item[{\bf Fig.~\ref{fig:plot1}}:]
Typical trajectory in the phase space plane $(R, P)$ for a repulsive potential $V$.
The trajectory comes from the separated fragments with a negative momentum $P_i$, slows down
due to the repulsive wall, touches a turning point (red dot) and follows a symmetric
path in the upper plane. An arrow indicates the direction of motion. Time $t = 0$ 
corresponds to passage at $R = R_i$, a value for which $V$ is vanishingly small.

\item[{\bf Fig.~\ref{fig:plot2}}:]
State-resolved absorption cross section in terms of the vibrational quantum number
$n$ of BC. The red and blue curves are found from Eqs. \eqref{3} and \eqref{17}
respectively.

\end{list}

\clearpage

\clearpage
\begin{figure}[!t]
\begin{center}
\scalebox{0.5}[0.5]{\rotatebox{0}{\includegraphics{plot2.eps}}}
\caption{\label{fig:plot2}}
\end{center}
\end{figure}

%
%
%
%
%
%
%
%
%
%
%
%
%
%
%
%
%
%

\end{document}